\documentclass[]{article}
\usepackage{graphicx}
\newcommand{\br}{{\bf r}}
\newcommand{\bk}{{\bf k}}
\newcommand{\bs}{{\bf s}}
\newcommand{\bq}{{\bf q}}

\begin{document}

\title{\bf Energy bands and Landau levels  of ultracold fermions in the bilayer honeycomb optical lattice}
\author{Jing-Min Hou\thanks{E-mail:
jmhou@seu.edu.cn}\\ {\em{Department of Physics, Southeast
University, Nanjing, 211189, China}} } \maketitle \

\begin{abstract} We investigate the spectrum and eigenstates of ultracold
fermionic atoms in the bilayer honeycomb optical lattice. In the low
energy approximation, the dispersion relation has parabolic form and
the quasiparticles are chiral.
  In the presence of the effective magnetic field,
which is created for the system with optical means, the energy
spectrum shows an unconventional Landau level structure.
Furthermore, the experimental detection of the spectrum is proposed
 with the Bragg scattering techniques.

\vspace{1cm} Keywords: optical lattices; ultracold atoms; Energy
bands; Landau levels.

\end{abstract}

\section{Introduction}
Recently, the studies of cold atoms in optical lattices are
extensively developed. Optical lattices are crystals made of light
periodic potentials that confine ultracold
atoms\cite{Jaksch,Greiner}.  Because of their precise control over
the system parameters  and defect-free properties, ultracold atoms
in optical lattices provide an ideal platform to study many
interesting physics in condensed matters\cite{Lewenstein}
 and even high energy physics\cite{Rapp}.

Very recently, a strong interest has been raised in the
two-dimensional honeycomb lattice \cite{Zhao,Zhu,Wu},  for its
physics  is closely related to that of the graphene
material\cite{Zheng,Novoselov,Li,Zhang,Jackiw,Hou}, which has
surprisingly rich collective behaviors. With the tight-binding
approximation, graphene has a linear dispersion relation resembling
 the Dirac spectrum for massless fermions.  In the
presence of a magnetic field, it has the Landau energy level with
square-root dependence on the quantum number $n$, instead of the
usual linear dependence. In particular, the zero-energy Landau level
exists at $n=0$, which is a direct result of chirality. Recently,
McCann {\it et al}. have studied the electronic states and
unconventional Landau levels of the bilayer graphene arranged
according to Bernal stacking\cite{McCann,McCann2}.

In this paper, we investigate the eigenstates  and spectrum of
ultracold fermions in the bilayer honeycomb optical lattice with a
different stacking order from that in Reference
\cite{McCann,McCann2}. In the absence of an effective magnetic
field, the dispersion relation has parabolic form and the
quasiparticles are still chiral like that in the monolayer system.
In the presence of   an effective magnetic field, which can be built
by coupling the internal states(spin) of atoms to spatially varying
laser beams \cite{Juzeliunas,Juzeliunas1,Juzeliunas3,Liu,Zhu2}, the
spectrum shows an unconventional Landau level  structure. The
experimental detection of the spectrum is proposed with  the Bragg
scattering techniques.

\section{The model}

\begin{figure}[ht]
 \includegraphics[width=1\columnwidth]{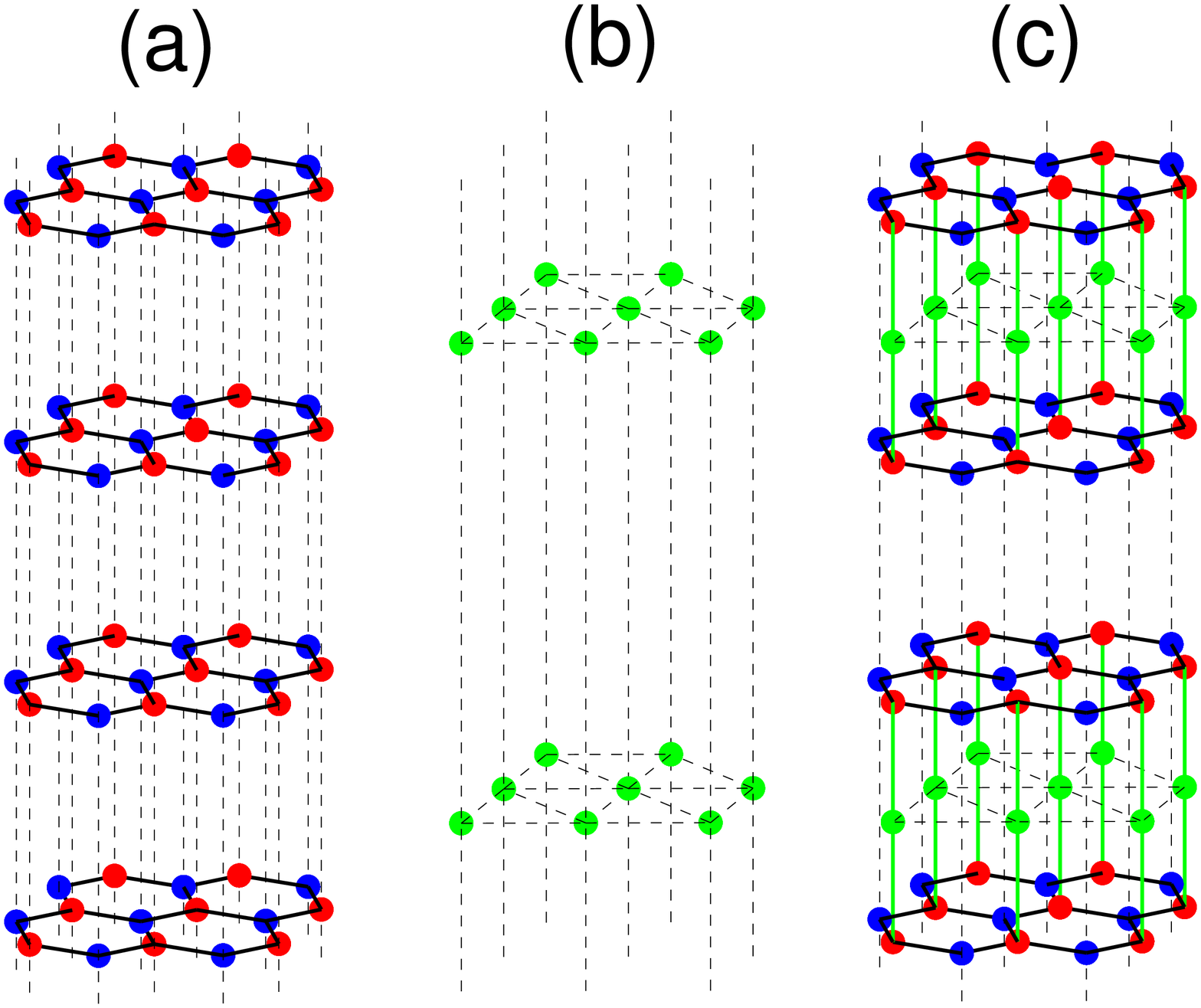}
\caption{(a) The independent monolayer honeycomb lattice. (b) The
adding triangular lattice. (c) The bilayer honeycomb lattice built
with putting (a) and (b) together.} \label{fig1}
\end{figure}

 We
consider a system of ultracold fermions confined in  the bilayer
honeycomb lattice. The honeycomb lattice consists of two sublattices
denoted by $A$ and $B$. Then, the bilayer honeycomb lattice
considered in our work  is formed by coupling the $B$ sublattices of
the two layers with tunneling and  leaving  the $A$ sublattices of
the two layers uncoupled. One can create the bilayer honeycomb
lattice in the following steps. First,  one builds the monolayer
honeycomb lattice as shown in Fig.\ref{fig1} (a)
 with three laser beams in the $x-y$ plane and two
laser beams along the $z$ direction\cite{Duan}.  When the potential
barrier of the optical lattice along the $z$ direction is high
enough, the vertical tunneling between different planes is
suppressed seriously, then every layer is an independent
two-dimensional honeycomb lattice. Secondly, one  makes the
triangular lattice as shown in Fig.\ref{fig1} (b) with red-detuning
laser fields\cite{Grynberg}. Finally, to realize the bilayer
honeycomb lattice, one can  put the triangular lattice and the
honeycomb lattice together as shown in Fig.\ref{fig1} (c). There
exists an additional micro-trap between sublattices $B$ of every two
layers in the honeycomb lattice. The additional micro-trap lowers
the barrier between sublattices $B$ of these two layers in the
honeycomb lattice, or links sublattices $B$ of these two layers in
the honeycomb lattice  as an intermediate point, so that sublattices
$B$ of these two layers in the honeycomb lattice  are coupled.
 Following this scheme, many independent bilayer honeycomb lattices
 can be
achieved (see Fig.\ref{fig1} (c)), so we only need to investigated
one of them. For convenience, we assume that the whole system is
trapped in a two-dimensional box, which can be achieved by adding
four blue detuning endcap beams at the edges of the optical lattice
in $x-y$ plane\cite{Meyrath}. With this box trap, the system can be
considered to have the hard wall boundary condition approximately,
so we can neglect the boundary effect in our discussion.

\begin{figure}[ht]
 \includegraphics[width=0.5\columnwidth]{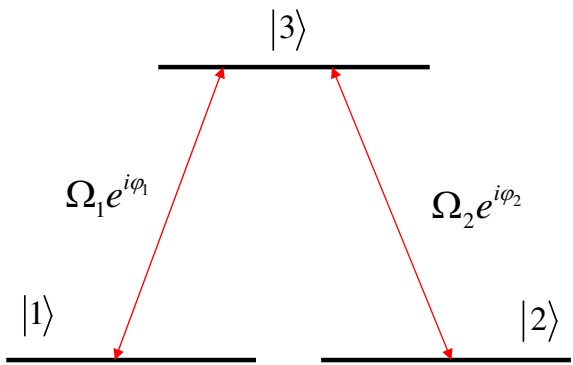}
\caption{The light-atom interactions between fermionic atoms and two
laser beams. } \label{fig2}
\end{figure}

  In this scheme, the ultracold atoms have a $\Lambda$-type
three-level configuration, the states $|1\rangle$ and $|2\rangle$
are degenerate states, which  are assumed to be different Zeeman
states on the same hyperfine level, and $|3\rangle$ is an excited
state. The  ground state $|j\rangle$ with $j=1,2$ and the excited
state $|3\rangle$ are coupled though two laser field with the
corresponding Rabi frequencies $\Omega_j e^{i\varphi_j}$,
respectively\cite{Juzeliunas3}.  The schematic representation of
this scheme is as shown in FIG. \ref{fig2}. The total Hamiltonian
reads, $\hat{H}=\hat{H_0}+\hat{H_1}$. The non-perturbative
Hamiltonian $\hat H_0$ is given by,
\begin{eqnarray}
 \hat{H}_0=\sum_\alpha
\int d\br
\hat\Psi^\dag_\alpha(\br)\left[-{\hbar^2}\nabla^2/{2m}+V(\br)\right]\hat\Psi_\alpha(\br),
\end{eqnarray}
 and the light-atom interaction Hamiltonian $\hat H_1$ is given by,
\begin{eqnarray}
\hat{H}_1=\int d\br \left[\Omega_1
e^{i\varphi(\br)}\hat\Psi_3^\dag(\br)\hat\Psi_1(\br)+\Omega_2\hat\Psi_3^\dag(\br)\hat\Psi_2(\br)+{\rm
H.c.}\right].
\end{eqnarray}
Diagonalizing the interaction Hamiltonian with  the  unitary
transformation $S$,
\begin{eqnarray}
S=\left( \begin{array}{ccc} \cos\theta& -\sin\theta e^{-i\varphi}&0
\cr\frac{\sqrt{2}}{2}\sin\theta
e^{i\varphi}&\frac{\sqrt{2}}{2}\cos\theta&-\frac{\sqrt{2}}{2}\cr
\frac{\sqrt{2}}{2}\sin\theta
e^{i\varphi}&\frac{\sqrt{2}}{2}\cos\theta&\frac{\sqrt{2}}{2}
\end{array}\right),
\end{eqnarray}
yields three eigenstates $|\Phi_1\rangle$, $|\Phi_2\rangle$ and $
|\Phi_3\rangle$, where $\tan\theta=|\Omega_1|/|\Omega_2|$ and
$\varphi=\varphi_1-\varphi_2$ are both  position-dependent
variables.
 The corresponding eigenvalues are $
E_i=(0,-\sqrt{|\Omega_1|^2+|\Omega_2|^2},\sqrt{|\Omega_1|^2+|\Omega_2|^2})
$. The new field operators corresponding to the eigenstates are
related with the old field operators as
\begin{eqnarray}
\left(
\begin{array}{c}\hat\Phi_1\\ \hat\Phi_2\\
\hat\Phi_3\end{array}\right)=S\left(\begin{array}{c} \hat\Psi_1\\
\hat\Psi_2\\ \hat\Psi_3\end{array}\right).\end{eqnarray} In the new
bases, and under the adiabatic condition $\langle \Phi_1|\hat
H_0|\Phi_j\rangle\ll |E_i-E_j|$ for $j=2,3$, we can apply the
adiabatic condition and then neglect the populations of the states
$|\Phi_2\rangle$ and $|\Phi_3\rangle$. Therefore, the effective
Hamiltonian can be rewritten in the dark-state basis
$|\Phi_1\rangle$
 \cite{Juzeliunas,Juzeliunas1,Juzeliunas3,Liu,Zhu2},
\begin{eqnarray}
\hat{H}=\int d\br
\hat\Phi_1^\dag(\br)\left[\frac{1}{2m}(-i\hbar\nabla-{\bf
{A}})^2+{V}_{eff}(\br)\right]\hat\Phi_1(\br), \label{eh}
\end{eqnarray}
where ${\bf A}=-\hbar\sin^2\theta \nabla\varphi$ and $\hat{\cal
H}\equiv\frac{1}{2m}(-i\hbar\nabla-{\bf {A}})^2+{V}_{eff}(\br)$ is
the single particle Hamiltonian with $V_{eff}(\br)$ being the
effective trap potential.
 Here $\bf
A$ is the effective gauge potential associated with the artificial
magnetic field ${\bf B}=\nabla\times {\bf A}$. In the practical
case, we choose two counter-propagating Gaussian laser beams as
 $\Omega_j e^{i\varphi_i}=\Omega_0 \exp[-(x-x_j)^2/\sigma_0^2]\exp(-ik_j y)$\ $(j=1,2)$, where the
 propagating wave vectors $k_1=-k_2=k_0/2$ and the center position $x_1=-x_2=\Delta x/2$
\cite{Juzeliunas3}. Then the effective trap potential is,
\cite{Juzeliunas3}
\begin{eqnarray} V_{eff}({\bf
r})=V({\bf r})+\frac{\hbar^2
k_0^2}{2m}\frac{(1+1/4d^2k_0^2)}{4\cosh^2(x/2d)},
\end{eqnarray}
and the effective vector gauge potential is
\begin{eqnarray} {\bf
A}=\frac{\hbar k_0}{1+e^{-x/d}}{\bf e}_y,\end{eqnarray}
 with
$d=\sigma_0^2/(4\Delta x)$. Straightforwardly, one can obtain the
effective magnetic field, \cite{Juzeliunas3}
\begin{eqnarray}
{\bf B}=\frac{\hbar k_0}{4d\cosh^2(x/2d)}{\bf e}_z.
\end{eqnarray}

Practically, one may set $d\sim1{\rm mm}$ and  $-0.01{\rm
mm}<x<0.01{\rm mm}$, so the condition $|x/d|\ll 1$ is satisfied.
 In this practical condition,
 the effective trap potential can approximately be
written as
\begin{eqnarray} V_{eff}({\bf
r})\approx V({\bf r})+\frac{\hbar^2
k_0^2}{2m}\frac{(1+1/4d^2k_0^2)}{4},
\end{eqnarray}
which has an additional constant term compared with the original
external trap potential. This additional constant term does not
change the geometrical structure of the original trap potential, so
we can drop out it as a constant chemical potential term.  The
effective magnetic field can approximately be written as
\begin{eqnarray}
{\bf B}\approx {\bf B}^{(0)}+{\bf B}^{(2)}=\frac{\hbar k_0}{4d}{\bf
e}_z -\frac{\hbar k_0}{4d} \frac{x^2}{8d^2}{\bf e}_z.
\end{eqnarray}
where the quadratic term can be neglected for
$|B^{(2)}/B|<1.25\times 10^{-5}$. Thus,
 the effective magnetic field can be regarded as a homogeneous one in the regime considered in our scheme.
 For the typical parameter value $k_0\sim 2\times 10^6 {\rm m}^{-1}$
 and $d\sim  10^{-3} {\rm m}$, we obtain the magnitude of the
 effective magnetic field, $B\sim 3.3\times 10^{-25}{\rm J\cdot
 s\cdot m^{-2}}$.

\section{The effective low energy Hamiltonian}

Taking the tight-binding limit, we can superpose the Bloch states to
get two sets of Wannier functions $w_\alpha^A(\br-\br_i)$ and
$w_\alpha^B(\br-\br_j)$ with $\alpha=1,2$, which correspond to
sublattices $A$ and $B$ of layer $\alpha$,  respectively. In the
presence of the effective gauge field we can expand the field
operator in the lowest band Wannier functions as, \begin{eqnarray}
\hat\Phi_1(\br)&=&\sum_{\alpha=1,2}\left[\sum_{i\in
A}\hat{a}_\alpha(\br_i)e^{\frac{i}{\hbar}\int_0^{{\bf r}_i}{\bf
A}\cdot d{\bf r}}w_A(\br-\br_i)\right.\nonumber\\&&\left.+\sum_{j\in
B}\hat{b}_\alpha(\br_j)e^{\frac{i}{\hbar}\int_0^{{\bf r}_j}{\bf
A}\cdot d{\bf r}}w_B(\br-\br_j)\right] \label{wannier}.
\end{eqnarray} Substituting the above expression into Eq.(\ref{eh}),
we can rewrite the Hamiltonian as $\hat{H}=\hat{H}_0+\hat{H}_1$ with
\cite{Jackiw},
\begin{eqnarray}
\hat{H}_0&=&-t\sum_\alpha\sum_{\br_i \in
A}\sum_{j=1,2,3}[\hat{a}_\alpha^\dag(\br_i)\hat{b}_\alpha(\br_i+\bs_j)e^{\frac{i}{\hbar}\int_0^{{\bf
s}_j}{\bf A}\cdot d{\bf r}}+{\rm H.c.}],\label{h0}
\end{eqnarray}
and
\begin{eqnarray}
\hat{H}_1&=&-t_\perp\sum_{\br_i \in
A}[\hat{b}_1^\dag(\br_i)\hat{b}_2(\br_i)+{\rm H.c.}],\label{h1}
\end{eqnarray}
 where $t$ is the tunneling parameter with $t=-\int d{\bf
r}w_A^*(\br-\br_i)\hat{\cal H}_0w_B(\br-\br_j) =-\int d{\bf
r}w_B^*(\br-\br_j)\hat{\cal H}_0w_A(\br-\br_i)$; $t_\perp=-\int
d{\bf r}w_{B_2}^*(\br-\br_i)\hat{\cal H}_0w_{B_1}(\br-\br_j)$; the
energy shifts for sublattice A and B are $\epsilon_A=\int d{\bf
r}w_A^*(\br-\br_i)\hat{\cal H}_0w_A(\br-\br_i)$ and $\epsilon_B
=\int d{\bf r}w_B^*(\br-\br_j)\hat{\cal H}_0w_B(\br-\br_j)$
respectively, with $\hat{\cal
H}_0\equiv-\frac{\hbar^2}{2m}\nabla^2+{V}_{eff}(\br)$. Here, for
convenience, we can have dropped out a constant  term in Hamiltonian
(\ref{h0}). The three vector $\bs_j$ in Eq.(\ref{h0}) are $
\bs_1=(0,-1)a,\ \bs_2=\left({\sqrt{3}}/{2},{1}/{2}\right)a,\ $ and
$\bs_3=\left(-{\sqrt{3}}/{2},{1}/{2}\right)a $, where $a$ is the
lattice spacing. The three vector $\bs_i (i=1,2,3)$ connect any site
of sublattice $A$ to its nearest neighbor sites belonging to
sublattice $B$ in every layer.

Here, we assume  the condition $t_\perp\ll t$ being satisfied,  so
that we can consider Eq. (\ref{h1}) as a perturbation.
 We take the
Fourier transformation to $\hat{a}(\br)$ and $\hat{b}(\br)$ as,
\begin{eqnarray}
&&\hat{a}_\alpha(\bk)=\sum_{{\bf r}_i\in A}e^{-i {\bf k}\cdot{\bf
r}_i} \hat{a}_\alpha(\br),\\ &&\hat{b}_\alpha(\bk)=\sum_{{\bf
r}_i\in B}e^{-i {\bf k}\cdot{\bf r}_i} \hat{b}_\alpha(\br_i),
\end{eqnarray}
where $\br $ is the coordinate on $x-y$ plane.
 Substituting the
above expressions into Eqs.(\ref{h0}) and (\ref{h1}), we obtain the
following Hamiltonian,
\begin{eqnarray}
&&\hat{H}_0=\sum_\alpha\sum_k
[\xi(\bk)\hat{a}_\alpha^\dag(\bk)\hat{b}_\alpha(\bk)+\xi^*(\bk)\hat{b}_\alpha^\dag(\bk)\hat{a}_\alpha(\bk)],\\
&&\hat{H}_1=-t_\perp\sum_{\bk}[\hat{b}_1(\bk)^\dag
\hat{b}_2(k)+\hat{b}_2(k)^\dag \hat{b}_1(k)],
\end{eqnarray}
 where $\xi(\bk)$ is the single-particle energy spectrum without interlayer tunneling  and
defined via  $ \xi(\bk)=-t\sum_{j=1,2,3}e^{-i(\bk\cdot
\bs_i-\frac{1}{\hbar}\int_0^{{\bf s}_j}{\bf A}\cdot d{\bf r})}$. The
energy spectrum contains two zero-energy points at ${\bf K}_\pm
=\pm({4\pi}/{3\sqrt{3}a},0) $ around which it is linearized.
Neglecting the coupling between the Fermi points ${\bf K}_\pm$, the
total Hamiltonian $\hat{H}$ can be expand around the contact point
${\bf K}_+ ({\bf K}_-)$ in coordinate space. Without loss of
generality,  we expand the total Hamiltonian
  around the contact point ${\bf K}_+$  as\cite{Jackiw,Hou},
\begin{eqnarray}
 \hat H=\int d^2r \hat\psi^\dag(\br)\hat{\cal
 H}\hat \psi(\br),
\end{eqnarray}
where the spinor $ \hat\psi=( {\hat\psi_1^a \ \ \hat\psi_1^b\ \
\hat\psi_2^a\ \ \hat\psi_2^b})^T $ for the Dirac point ${\bf K}_+$.
Here,  $\hat{\cal H}$ takes the $4\times 4$ matrix form,
\begin{eqnarray}
\hat{\cal H}=\hbar\left(\begin{array}{cccc} 0&v_F\hat\pi^\dag&0&0\cr
v_F\hat\pi&0&0&-t_\perp\cr 0&0&0&v_F\hat\pi^\dag\cr
0&-t_\perp&v_F\hat\pi&0 \end{array}\right),
\end{eqnarray}
 where
$\hat\pi=\hat{\pi}_x+i\hat{\pi}_y$ and
$\hat\pi^\dag=\hat{\pi}_x-i\hat{\pi}_y$, with
$\hat{{\pi}}_x=\hat{p}_x-{A}_x/\hbar$  and
$\hat{{\pi}}_y=\hat{p}_y-{A}_y/\hbar$, and $v_F=3at/2\hbar$ is the
Fermi velocity.
 Here, $t_\perp\ll t$ is assumed.
Eliminating the dimer state components $\hat\psi_1^b$ and
$\hat\psi_2^b$, we can reach a two-component Hamiltonian describing
effective hopping between the $A_1$-$A_2$ sites
\begin{eqnarray}
\hat{\cal H}_{\rm
eff}=\frac{\hbar^2}{2m}\left(\begin{array}{cc}0&\hat\pi^\dag\hat\pi\cr
\hat\pi^\dag\hat\pi&0\end{array}\right),\label{hk2}
\end{eqnarray}
where $m=t_\perp/2v_F^2$.

\section{Energy bands}

First, we consider the case without gauge fields, i.e.
$\hat{{\pi}}_x=-i\partial_x$  and $\hat{{\pi}}_y=-i\partial_y$.
The
eigenfunctions of the Hamiltonian (\ref{hk2}) are given by
\begin{eqnarray}
f_{s\bk}(\br)=\frac{1}{\sqrt{2}L}\exp(i\bk\cdot\br)\left(\begin{array}{c}s
\cr 1\end{array}\right),\label{eigen}
\end{eqnarray} where $L^2$ is the area of the
system, and $s$ denotes the conduction band with $s=+1$ and the
valence band with $s=-1$. The corresponding eigenenergies are
\begin{eqnarray}
E= \frac{s\hbar^2k^2}{2m},
\end{eqnarray}
where $k=\sqrt{k_x^2+k_y^2}$. This dispersion relation has parabolic
form  as shown in Fig.\ref{fig3}. Here, the quasiparticles are still
chiral like that in the monolayer system. The pseudospin vector
${\bf n}=(1, 0)$ is a constant for any wave vector $\bk$ in our
work, while ${\bf n}=(\cos(2\phi), \sin(2\phi))$ for
$\bk=(k\cos\phi, k\sin\phi)$  in the bilayer graphene with Bernal
stacking order as in References \cite{McCann, McCann2}. Thus, in our
bilayer honeycomb lattice configuration, the Berry phase $2\pi$ in
the bilayer
 graphene with Bernal stacking order  is absent.

\section{Unconventional Landau levels} For the  case with
 an effective magnetic field in the Landau gauge $(0, Bx, 0)$,
 the eigenfunctions of the
Hamiltonian (\ref{hk2})  can be obtained as,
\begin{eqnarray}
F_{nk_y}(\br)&=&\frac{1}{\sqrt{2L}}\exp(ik_y
y)\left(\begin{array}{c}\phi_{|n|}\cr{\rm sgn}(n)\phi_{|n|}
\end{array}\right),
\end{eqnarray}
with ${\rm sgn}(n)=(1,0,-1)$ for $(n>0,n=0,n<0)$ respectively, for
$n\neq 0$, and
\begin{eqnarray}
F_{0k_y}(\br)&=&\frac{1}{\sqrt{2L}}\exp(ik_y
y)\left(\begin{array}{c}\phi_{0}\cr \pm\phi_{0}\end{array}\right),
\end{eqnarray}
for $n=0$. Here, $\phi_n$ are harmonic oscillator eigenstates
 as
\begin{eqnarray}
\phi_{|n|}=\frac{1}{\sqrt{2^{|n|}|n|!\sqrt{\pi}u}}\exp\left[-\frac{1}{2}\left(\frac{x-u^2k}{u}\right)^2\right]
H_{|n|}\left(\frac{x-u^2k}{u}\right),
\end{eqnarray}
where the quantum nubmer $n$ is an integer and $u=\sqrt{\hbar/B}$.
The corresponding Landau energy levels are $E_n= n\hbar \omega_c $
with $n=\cdots,-2,-1,0,1,2,\cdots$ and $\omega_c=2v_F^2B/t_\perp$,
which is shown in Fig.\ref{fig4}. This spectrum has a linear form
like the conventional Landau level spectrum. However, the
quasiparticles are chiral in our scheme. The zero modes exist and
the zero energy level is twofold degenerate compared to the non-zero
energy levels.

To give a numerical evaluation, the typical values of the parameters
can be taken as $t\sim 10^{-30}{\rm J}$, $t_\perp\sim 10^{-33}{\rm
J}$, $a\sim 200{\rm nm}$, $B\sim 3.3\times 10^{-25}{\rm J\cdot
 s\cdot m^{-2}}$. We can then
estimate the magnitude of the cyclotron frequency
$\omega_c=5.7\times 10^{3}{\rm s^{-1}}$ and obtain the first gap of
Landau level for the monolayer honeycomb lattice
 $\Delta \sim 6\times 10^{-31} {\rm J}$.
 The temperature required to keep atoms in the zeroth Landau level
 is $43\ {\rm nK}$.
\section{Bragg spectroscopy}

 It is not easy to measure the Hall
conductivity of cold fermionic atoms in the bilayer honeycomb
lattice. However, an available method to detect the unconventional
Landau levels of ultracold fermions on the bilayer honeycomb lattice
is the Bragg spectroscopy\cite{Stamper-Kurn}, which is extensively
used to probe the excitation spectrum in condensed matter physics.
In the Bragg scattering, the atomic gas is exposed to two laser
beams, with wavevectors $\bk_1$ and $\bk_2$ and a frequency
difference $\omega$. The light-atom interaction Hamiltonian can be
written as,
\begin{eqnarray}
\hat H_B=\sum_{s_1, s_2,\bk, \bq}\Omega_B e^{i\bq\cdot\br}| f_{s_2,
\bk+\bq}\rangle\langle f_{s_1,\bk}|+{\rm H.c.},
\end{eqnarray}
where $s_i=1,2$.
 In our case, we consider the case of half filling, i.e., the
bands with $n\leq 0$ are fully occupied and the bands with $n>0$ are
empty. The half filling state can be prepared with the coherent
filtering scheme proposed in Reference \cite{Rabl}.  From the
Fermi's golden rule, we obtain the dynamic structure factor as
follows,
\begin{eqnarray}
S(\bq,\omega)&=&\frac{1}{N\hbar^2\Omega^2}\sum_{\alpha}|\langle
\phi_{\beta}^{(f)}|H_B|\phi_{\alpha}^{(i)}\rangle|^2\nonumber\\
&&\times\delta(\hbar\omega-E_{\beta}+E_{\alpha})
\end{eqnarray}
where $N$ is the total number of atoms in the system;
$\phi_{\alpha}^{(i)}$ denotes the initial state and
$\phi_{\alpha}^{(f)}$ denotes the final state; $\alpha$ represents
all quantum parameters of quantum state.

\begin{figure}[ht]
 \includegraphics[width=0.5\columnwidth]{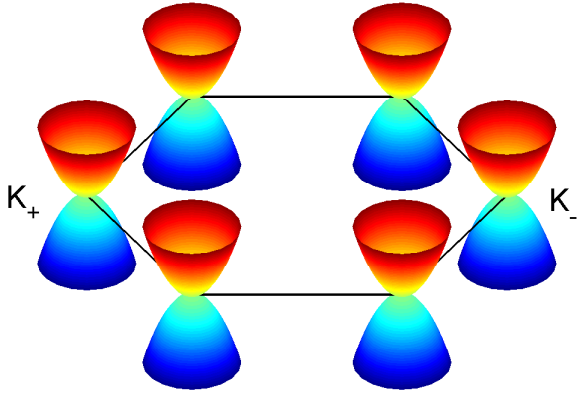}
\caption{ Energy bands of cold fermionic atoms in the bilayer
honeycomb lattice without an effective magnetic field.} \label{fig3}
\end{figure}

\begin{figure}[ht]
 \includegraphics[width=0.5\columnwidth]{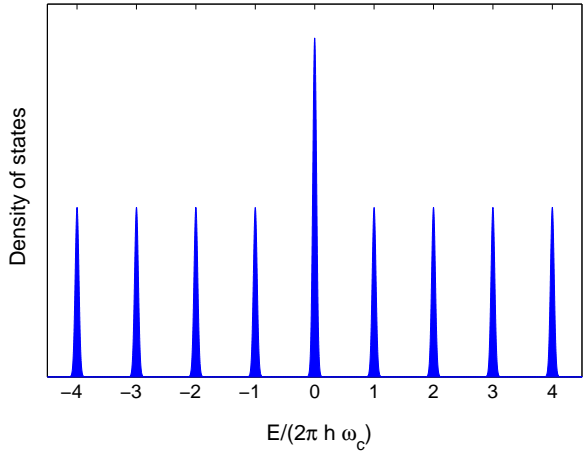}
\caption{ Landau levels  of cold fermionic atoms in the bilayer
honeycomb lattice with an effective magnetic field.} \label{fig4}
\end{figure}

\begin{figure}[ht]
\includegraphics[width=0.5\columnwidth]{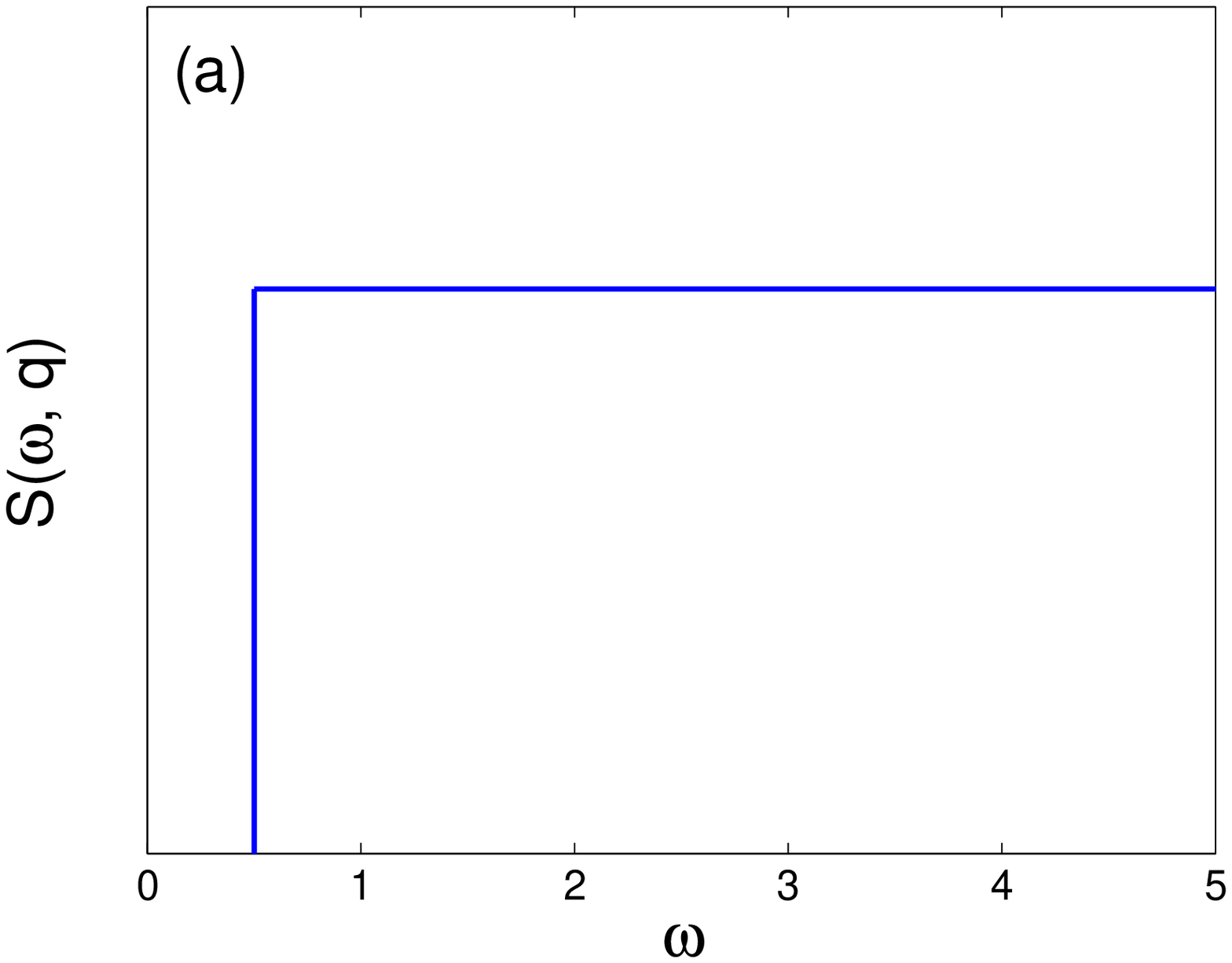}
\includegraphics[width=0.5\columnwidth]{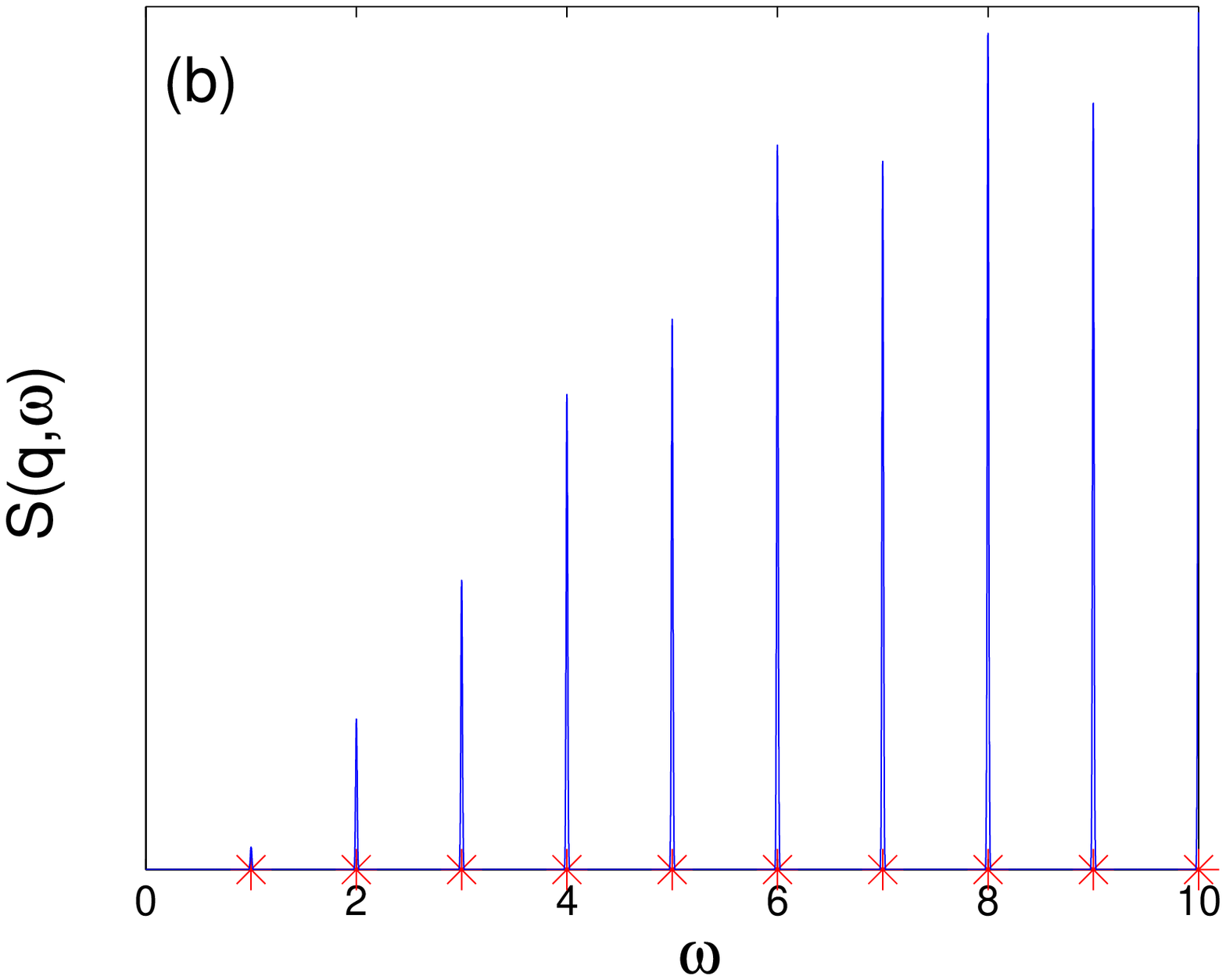}
\caption{The dynamic structure factors $S(q,\omega)$
 (not scaled) for  cold atoms in the bilayer honeycomb lattice  in the cases
 (a) without an effective magnetic field and (b) with an effective
 field. Here, $\omega$ is in units of $9at^2q^2/2\hbar t_\perp$ with $q=1.0\sqrt{B/\hbar}$.  } \label{fig5}
\end{figure}

 For simplicity, we assume that
the direction of $\bq$ is the same as the one of  $y$ axis in the
wavevector space, i.e. $\bq=q{\bf e}_y$. Following the above
formulae, we can straightforwardly evaluate the dynamic structure
factor $S(\bq,\omega)$.  Fig.\ref{fig5}(a) shows the dynamic
structure factor $S(\bq, \omega)$ as a function of $\omega$ for the
case without an effective magnetic field. We can find that $S(\bq,
\omega)$ is zero when $\omega$ is under $\hbar^2q^2/2m$ and is
finite constant for $\omega$ above $\hbar^2q^2/2m$. Fig.\ref{fig5}
(b) shows the dynamic structure factor $S(\bq, \omega)$ as a
function of $\omega$ with $q=1.0\sqrt{B/\hbar}$ for the bilayer
honeycomb lattice.  Similarly, when a zero-level atom is excited to
other states, the peaks are obtained at $\omega=(1, 2, \cdots,n,
\cdots)\Lambda^2 t^2/\hbar t_\perp $ with $\Lambda=
3a\sqrt{B/2\hbar}$, which are marked with red stars in
FIG.\ref{fig3} (b). The distances between the neighbor peaks marked
with red stars in Fig.\ref{fig3} (b) are identical.

\section{Conclusion}

 In summary, we have proposed a scheme to investigate
ultracold fermionic atoms in the  bilayer honeycomb lattice for the
cases without and with an effective magnetic field. The effective
magnetic  field can be built with optical techniques. For the case
without an effective magnetic field, the dispersion relation has
parabolic form  and the quasiparticles are   chiral. For the case
with an effective magnetic field, there exist unconventional Landau
levels that include a zero-mode level.
  We have calculated the dynamic structure factors for the two
cases and proposed to detect them with  the Bragg spectroscopy.

\section*{Acknowledgements}This work was  supported  by the
Teaching and Research Foundation for the Outstanding Young Faculty
of Southeast University.



\end{document}